# DISCOVERY OF THE NEUTRON RADIATIVE DECAY.

R.U. Khafizov [a], N. Severijns [b], O. Zimmer [c], H.-F. Wirth [c], D. Rich [c],
S.V. Tolokonnikov [a], V.A. Solovei [d], M.R. Kolhidashvili [d]

[a] RRC Kurchatov Institute, 123182 Moscow, Russia
[b] Katholieke Universiteit Leuven, B-3001 Leuven, Belgium
[c] Technische Universitat Munchen, 85747 Garching, Germany
[d] Petersburg Nuclear Physics Institute, 188350 Gatchina, Russia

Abstract

The aim of this work is experimental discovery and research of the neutron radiative beta-decay, where a new particle, the radiative gamma-quantum, is formed along with the expected decay products beta-electron, recoil proton and antineutrino. The discovery of this rare neutron decay mode was conducted through identification of triple coincidences events: simultaneous registration of beta electron, proton and radiative gamma-quantum. The ordinary neutron decay was registered by double coincidences of beta electron and recoil proton. The ratio of triple to double coincidences is connected with relative intensity of radiative neutron decay (branching ratio - B.R.) B.R. = $(3.2\pm1.6)\cdot10^{-3}$ (with 99.7 % C.L. and in the gamma energy region greater than 35 keV ), which we are the first in the world to measure during the second and third cycles on FRMII ( TUM, Germany ) in Summer 2005 [1]. This value of B.R. is consistent with standard electroweak theory.

**Introduction**.

It is well known that a radiative decay mode is expected for all elementary particle decays with charged particles in the final state. This decay branch is well established and was investigated for practically every elementary particle. However, the radiative decay of the free neutron

$$n \rightarrow p + e + \bar{\nu} + \gamma$$

is yet to be researched. The only exception is the experiment we conducted in 2002 on the intensive cold neutron beams at ILL (Grenoble, France), where for the first time in the world we received a limit for the relative intensity of this rare decay branch: B.R. < $6.9*10^{-3}$ ( 90% C.L.) [2]. This value exceeds the value we calculated and published only in by a few times. This fact, in turn, means that in our experiment of 2002 we came very close to discovering the radiative mode of neutron decay. In 2004 this result was internationally recognized in the famous review Particle Data Group [3].

Calculations of the neutron radiative spectrum in the framework of standard electroweak theory were carried out several years ago [4]. The calculated branching ratio for this decay mode as a function of the gamma energy threshold is shown in Figs. 1 and 2. The branching ratio for the energy region investigated here, i.e. over 35 keV, was calculated to be about $2\cdot10^{-3}$ ( gamma energy threshold ω on Fig. 1 and 2 is equal to 35 keV [4] ).

Given this rather large branching ratio of about two per thousand, it is in principle not a difficult task to measure it. In practice, however, a rather significant background, mainly caused by external bremsstrahlung emitted by the decay electrons when stopped in the electron detector, has to be overcome. In this experiment this was achieved with the help of a triple coincidence requirement between the electron, the gamma-quantum and the recoil proton. The presence of such a coincidence is used to identify a radiative neutron decay event, whereas an ordinary neutron beta decay is defined by the coincidence of an electron with a recoil proton. The latter coincidence scheme is routinely used to measure the emission asymmetry of electrons in the decay of polarized neutrons The set-up used in several of those experiments [5,6] was upgraded for the experiment on radiative neutron decay discussed here.

This upgrade mainly consisted of adding a gamma-detector into the existing vacuum chamber. However, the simple addition of a gamma-detector to the two already present detectors (for the decay electrons and recoil protons) would not suffice by itself. Indeed, in this experiment, besides the non-correlated background one also has to deal with a correlated background of bremstrahlung gamma-quanta that fully simulates the desired fundamental radiative decay process searched for. It was therefore necessary to consider this problem in more detail. This correlated background is caused by bremsstrahlung emission of the electron traveling through the electron detector and is quite significant even when the thickness of this detector is limited to only a few mm. It cannot be eliminated by requiring a triple coincidence of the electron, the photon and the proton. However, calculations [4] show that the radiative emission of a photon in neutron decay is not in the forward direction with respect to the electron emission direction, as in the case of bremsstrahlung, but reaches a maximum intensity at an angle of 35° (Fig. 3). It was this property of radiative neutron decay that led us to construct a segmented electron-gamma detector with a 35° angle between the sections for electron and gamma detection to reduce this background.

**2. Experimental set-up**

The experimental set-up is shown schematically in fig. 4. The intense cold neutron beam passes through a rather long neutron guide in which is installed a collimation system made of LiF diaphragms, placed at regular distances of 1 meter. The neutrons enter the vacuum chamber (1) through the last diaphragm (9) that is located directly before the decay zone. This zone is observed by three types of detector: the micro channel plate (MCP) proton detector (3), the electron detector (14) consisting of a 7 cm diameter and 3 mm thick plastic scintillator, and six gamma detectors (11) that are located on a ring centered around the electron detector and which consist of photomultiplier tubes each covered with a layer of CsI(Tl) scintillator. The thickness of these 7 cm diameter CsI(Tl) scintillators is 4 mm and has been selected so as to have a 100% detection efficiency for photons with energies over 100 keV. The six gamma detectors (11) surround the electron detector (13) (cf. the lower part of fig. 4) at an angle of 35° and are shielded from it by 6 mm of lead (12). By requiring a coincidence between the electron detector and any of the gamma detectors the bremsstrahlung background can in principle be overcome completely, because bremsstrahlung emission occurs only in the section that registers the electron. In this case, part of the statistics is lost, of course, as can be seen from figure 3. However, the neutron beam intensity of $10^{12}$ n/s in our experimental chamber is sufficiently enough to compensate for that loss and still allows for a good count rate. Recoil protons, formed in the decay zone, pass through a cylindrical time of flight electrode (7) in the direction of the proton detector (3) and are focused onto this detector with the help of spherical focusing electrodes (2). The focusing electrostatic field between the high voltage spherical and cylindrical electrodes (2) and (7) is created by the grids (5) and (6) at one side and by the proton detector grid (4), at ground potential, at the other side. It is important to note that the recoil protons take off isotropic from the decay point. In order not to loose half of the protons emitted, an additional grid (10) is added on the other side of the decay volume. The potential difference between the grid (10) and the grids and electrodes at the other side of the decay volume in principle assures a 4π solid angle coverage for the recoil protons.

The start signal that opens the time windows for all detectors is the signal from the electron, registered in the electron detector (13). For an event to be considered as a radiative neutron decay event there have to be simultaneous signals from the electron detector (13) and one of the gamma detectors (11), followed by a delayed signal from the proton detector (3). It is important to note that in the case of radiative decay, the gamma quantum in our equipment is registered by gamma detectors (11) surrounding the electron detector (13) before the electron is registered by the electron detector. In other words, electron is delayed in comparison to the radiative gamma quantum. In the future namely this fact will allow us to distinguish the peak of radiative gamma quanta in the triple coincidences spectrum. Besides these triple coincidences also double electron-proton coincidences, signaling an ordinary neutron decay event, are monitored.

It is important to note here that thanks to the LiF ceramics diaphragm system which was installed in the neutron beam line, the gamma background from the intense cold neutron beam was significantly suppressed. The background level in the gamma detector amounted to about 2.5 kHz only (at a neutron beam intensity of $10^{10}$ n/s). If the number of the diaphragms in the neutron guide were doubled, the background of the gamma detectors could be further reduced by another order of magnitude, thus becoming comparable to the noise of the photomultiplier tubes. Another important note is that in our last experiment at ILL we succeeded at obtaining a gamma background that was smaller by an order. This could be explained by the fact that on the "Mephisto" beam at FRMII we used a collimation system that was reduced in comparison to the original. Besides, our experiment was the first to be conducted on the intensive cold beam neutron "Mephisto" and as it turned out the axis of this beam went a little higher than the axis of our collimation system, which also contributed to a significant increase in the gamma background. The count rate in the electron detector was just about 100 Hz. It is very likely that most of this count rate is due to electrons from neutron decay since the count rate in this detector almost immediately dropped to zero when the neutron beam was switched off. The main problem in this experiment was the proton detector background, which turned out to be very sensitive to the vacuum conditions in the experimental chamber.

## 3. Results

All components of the experimental set-up performed well. Electron-proton coincidences could clearly be observed, while the sectioned electron-gamma detector performed as expected. The results obtained with a vacuum less than $10^{-6}$ mbar, are presented in figures 5 and 6. Figure 5 shows the electron-proton coincidence spectrum, while the corresponding triple electron-proton-gamma coincidence spectrum is shown in figure 6. Results of the experiment on radiative neutron decay are presented on Fig. 5 and 6. Fig. 5 presents the summary statistics on double e-p coincidences (coincidences of electron with delayed proton), and Fig. 6 presents the summary statistics on triple e-p-γ coincidences (coincidences of electron, gamma-quantum, and delayed proton). Fig. 5 clearly shows two major peaks: one peak with a maximum in channel 99, which is the peak of zero or false [6] coincidences. The position of this peak marks the zero time count, namely the time when the electron detector registered the electron. The next peak visible on Fig. 5 has a maximum in channel 120, and is the peak of e-p coincidences of electron with delayed proton. Analogous situation was observed in experiments on the measurement of the correlation coefficients [6]. Fig. 5 shows that the total number in e-p coincidences peak in our experiment equals $N_D=3.75 \cdot 10^5$. This value exceeds the value we obtained in our previous experiment conducted on beam PF2 at ILL by two orders. It was precisely because of the low statistics volume that we could not then identify the events of radiative neutron decay, and defined only the upper B.R. limit [2].

Fig. 6 of triple coincidences clearly shows three peaks, and the leftmost peak with the maximum in channel 102 is connected to the peak of the radiative gamma-quanta in question, as this gamma-quantum is registered by the gamma detectors in our equipment before the electron. Comparing Fig. 5 and 6, it becomes clear that if we ignore the third leftmost peak with the maximum in channel 102 in Fig. 6, the spectrum of double e-p coincidences will resemble the spectrum of triple e-p-γ coincidences. The peak with the maximum in channel 106 on Fig. 6 is connected to the left peak of false coincidences on Fig. 5, and the peak with the maximum in channel 120 on Fig. 6 is connected to the right peak of e-p coincidences on Fig. 5. The emerging picture becomes obvious when one uses a standard procedure, introducing a response function for gamma channel $R_\gamma(t,t')$, which is necessary also for calculating the number of triple radiative coincidences $N_T$ in radiative peak.

Now we can define the response function of gamma channel $R_\gamma(t,t')$ in the standard way:

$$S_{out}(t)=\int S_{in}(t')R_\gamma(t,t')dt' \qquad (1)$$

In our case, $S_{in}(t)=S_D(t)$ is the spectrum of double e-p coincidences on Fig. 5, while $S_{out}(t)$ is the background spectrum of triple coincidences. In a situation with a perfect detector, the response function is local one $R(t,t')=k\delta(t-t')$. From this it follows that the spectrum $S_{out}(t)$ will resemble $S_{in}(t)$ and a comparison of Fig. 5 and 6 would allow us to define coefficient k=0.018. Here one must note that the value of this coefficient is proportional to the number of the background gamma-quanta. In our case the average number of gamma quanta in all 6 channels was 15000 γ/s, and if we can reduce that value by an order (as we had done in the previous experiment conducted at ILL [2]), the value of k will also go down by an order.

However, for real detectors the response function is always not local, which leads to a deformation of spectrum $S_{out}(t)$. In our case this non-local behavior is connected to the fact that the duration of pulse front from gamma detectors is on average 200 ns (which equals to 8 time channels on Fig. 5 and 6), and the duration of pulse front from the electron detector equals to 10 ns. This difference explains, in particular, the shift of the peaks on Fig. 6: the peak with maximum in channel 105, connected to the peak of false coincidences, is shifted to the right (the side of delay) in comparison to the peak of false coincidences in channel 99 on Fig. 5. After analyzing the spectra with the help of the non-local response function following formula (1), we finalize the average value for the number of radiative neutron decays $N_T=360$ with a statistics fluctuation of 60 events.

The triple coincidence count rate $N_T$ can be expressed through B.R. as

$$N_T = \frac{N_D}{\varepsilon_e \, \Omega_e \, \varepsilon_p \, \Omega_p} \, \varepsilon_e \, \Omega_e \, \varepsilon_p \, \Omega_p \, \varepsilon_\gamma \, \Omega_\gamma \, f \, BR \quad (2)$$

where $N_D$ is the e-p coincidence count rate and $\varepsilon_i$ and $\Omega_i$ (i = e, p, γ) are respectively the efficiencies and the solid angles for the electron detector, the proton detector and the six gamma detectors. Further, the product $\Omega_\gamma f$ stands for the integral of the normalized photon-distribution function $f$ (which reaches a maximum at 35°; cf. fig. 3) over the stereometric angle of the six gamma-detectors and $BR$ is the branching ratio of the radiative decay mode for the observed energy region. Finally eqn. (2) changes to

$$BR = \frac{N_T}{N_D} \, (\varepsilon_\gamma \, \Omega_\gamma \, f)^{-1} \quad (3)$$

Note that this result is independent of the efficiency or solid angle of both the electron and the proton detectors. With the number of observed double e-p coincidences $N_D = 3.75 \cdot 10^5$, triple e-p-γ coincidences $N_T = 360$, $\varepsilon_\gamma = 1$ and $\Omega_\gamma f = 0.3$ one then deduces the value for radiative decay branching ratio of $(3.2 \pm 1.6) \cdot 10^{-3}$ (99.7 % C.L.) with the threshold gamma energy ω=35 keV .

## 4. Conclusion

Results from the first experiment aiming to observe the as yet undiscovered radiative decay mode of the free neutron are reported. Although the experiment could not be performed under ideal conditions, the data still allowed one to deduce the B.R. = $(3.2 \pm 1.6) \cdot 10^{-3}$ (99.7 % C.L.) for the branching ratio of radiative neutron decay in the gamma energy region greater than 35 keV. This value is in agreement with the theoretical prediction based on the standard model of weak interactions.

Taking into account the fact that the experimental conditions can still be significantly optimized, an e-p coincidence count rate of 5-10 events per second is within reach. Together with the standard model prediction for the branching ratio of this decay mode, this would correspond to a triple e-p-γ coincidence rate of several events per 100 seconds. This can easily be observed with the current experimental set-up, which is now being optimized with a view to performing such an experiment. The aim of that experiment will then not only be to establish the existence of radiative neutron beta decay, but also to study the radiative gamma spectrum in more detail. This, in turn, would allow to discover the deviation from standard electroweak theory. The average B.R. value we obtained deviates from the standard model, but because of the presence of a significant error (50%) we cannot make any definite conclusions. Precision level must be increased. According to our estimates, we will be able to make more definite conclusions about deviation from the standard electroweak theory at the precision level of less than 10%.

**Acknowledgement**

The authors would like to thank Profs. D. Dubbers, J. Deutsch and Drs. T. Soldner, G. Petzoldt, I. Konorov and S. Mironov for valuable remarks and discussions. We are also grateful to the administration of the FRMII, especially Profs. K. Schreckenbach and W. Petry for organizing our work. We would especially like to thank RRC President Academician E.P. Velikhov and RRC Director I.N. Polyakov for their support, without which we would not have been able to conduct this experiment. Financial support for this work was obtained from INTAS (Project N 1-A −115; Open 2000), the RFBR (Project N 03-02-16261) and the Flemish Fund for Scientific Research (F.W.O.).

# Figure captions

**Fig. 1**
The expected standard model branching ratio for radiative neutron beta decay (summed over all energies larger than the threshold energy ω) as a function of ω (from [3]).

**Fig. 2**
The expected standard model branching ratio for radiative neutron beta decay (summed over all energies larger than the threshold energy ω) as a function of ω, for the low energy part of the spectrum (from [3]).

**Fig. 3**
Dependence of the radiative decay spectrum on the angle Ξ between the photon and the electron momenta (upper curve for a threshold gamma energy of 25 keV, lower curve for a threshold gamma energy of 50 keV) (from [3]).

**Fig. 4**
Schematic lay-out of the expemrimental set-up.
(1) detector vacuum chamber, (2) spherical electrodes to focus the recoil protons on the (at 18-20 kV), (3) proton detector, (4) grid for proton detector (at ground potential), (5) & (6) grids for time of flight electrode, (7) time of flight electrode (at 18-20 kV), (8) plastic collimator (5 mm thick, diameter 70 mm) for beta-electrons, (9) LiF diafragms, (10) grid to turn the recoil proton backward (at 22-26 kV), (11) six photomultiplier tubes for the CsI(Tl) gamma detectors, (12) lead cup, (13) photomultiplier tube for the plastic scintillator electron detector.

**Fig. 5**
Time dependent spectrum for e-p coincidences. Each channel corresponds to 25 ns. The peak at channel 99 corresponds to the prompt coincidences. The coincidences between the decay electrons and delayed recoil protons (e-p coincidences) are contained in the large peak centered at channel 120.

**Fig. 6**
Time dependent spectrum for triple e-p-g coincidences. Each channel corresponds to 25 ns. In this spectrum, three main peaks in channels 103, 106 and 120 can be distinguished. The leftmost peak in 103 channel among these three main peaks is connected with the peak of radiative decay events.

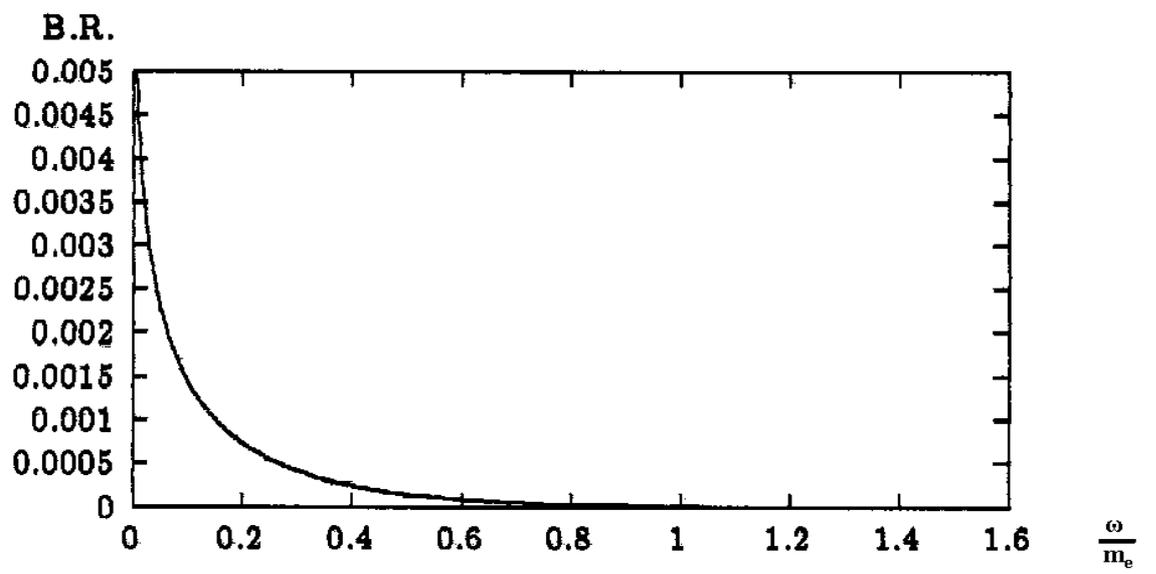

**Fig. 1**

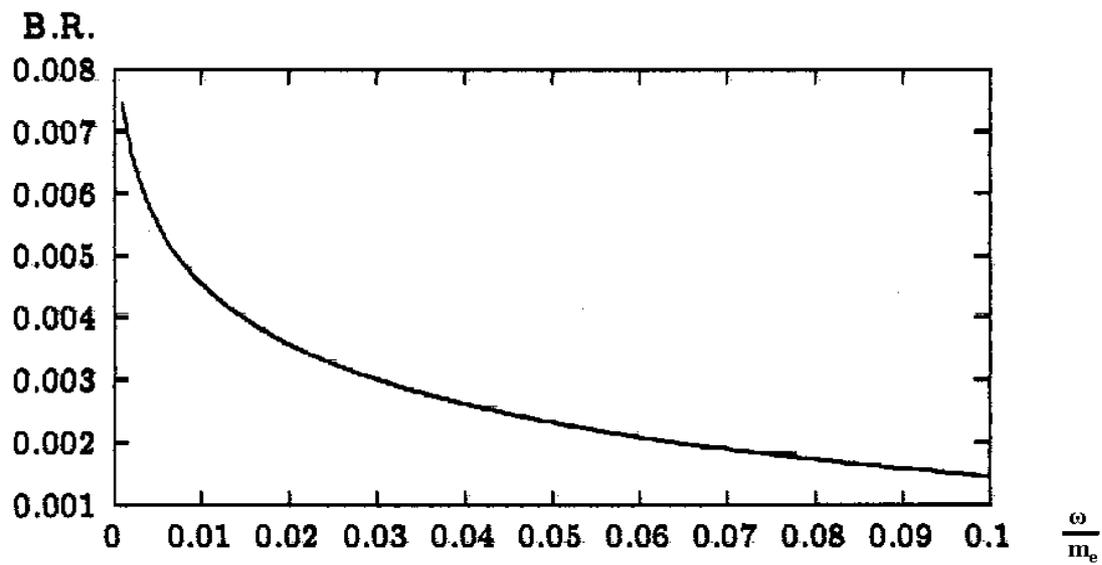

Fig. 2

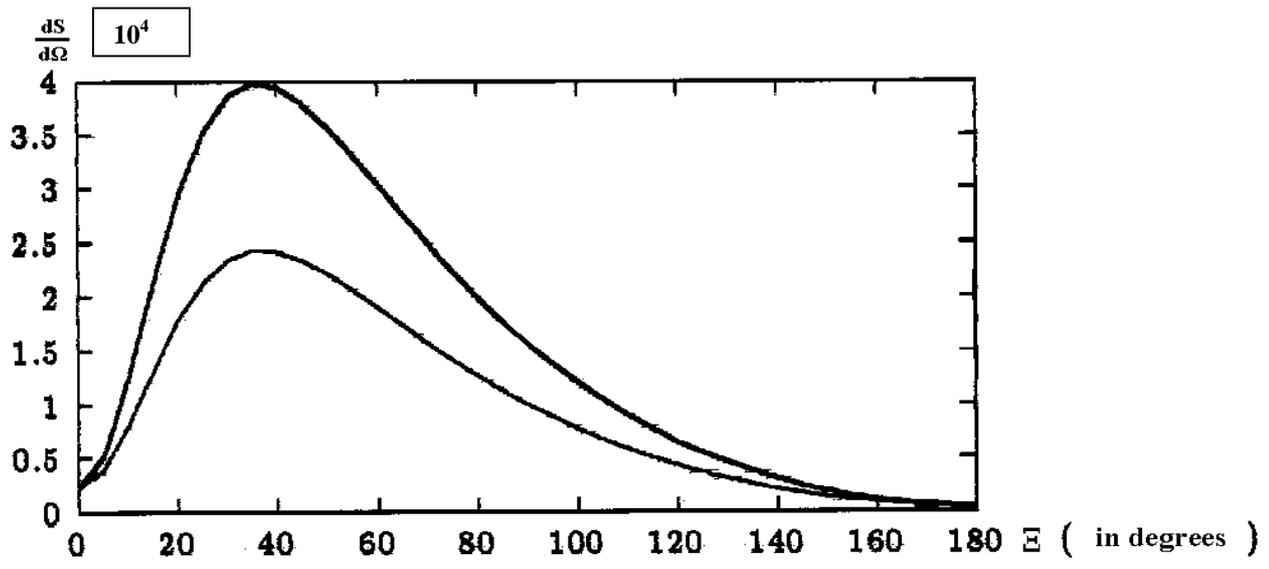

**Fig. 3**

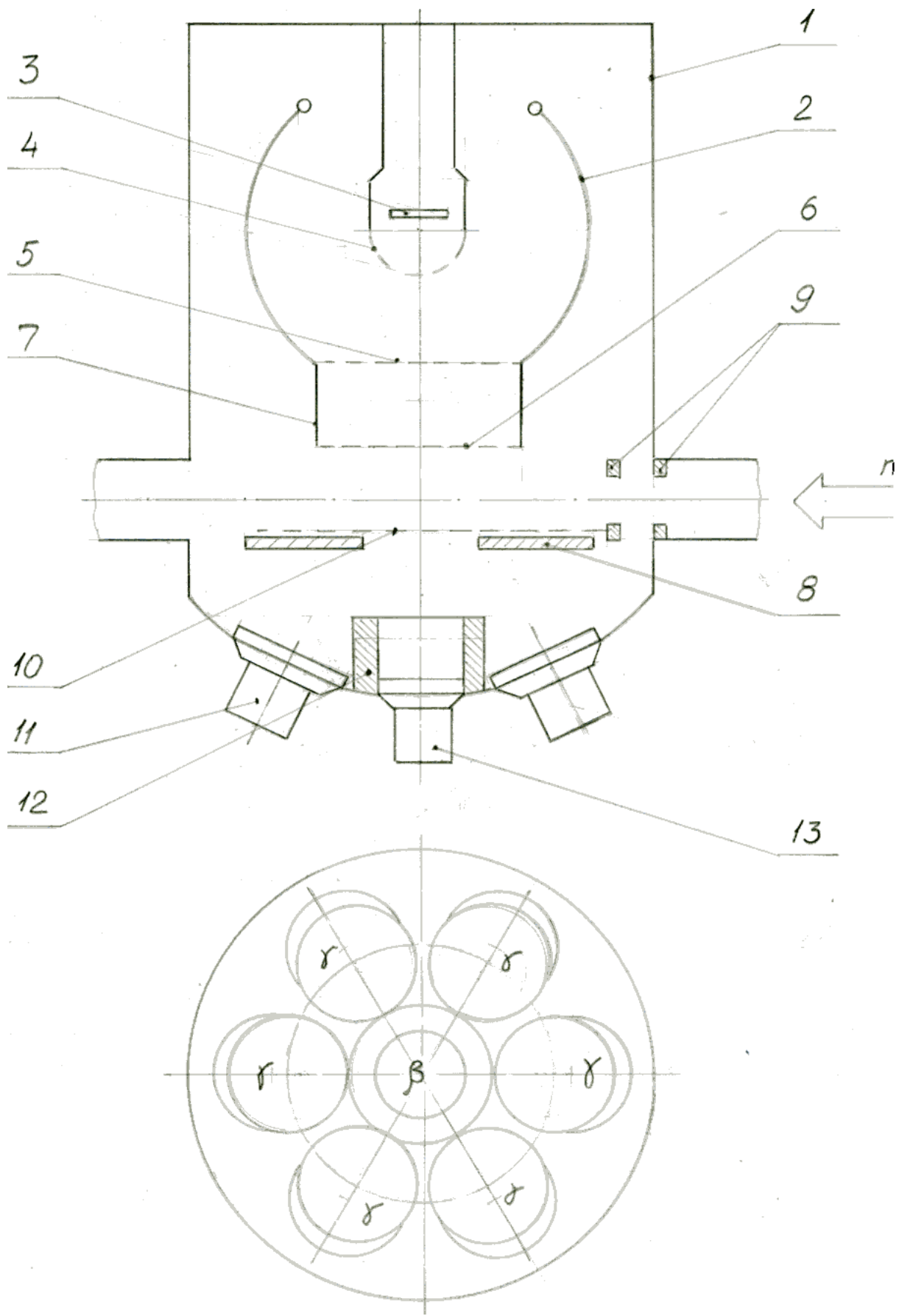

**Fig. 4**

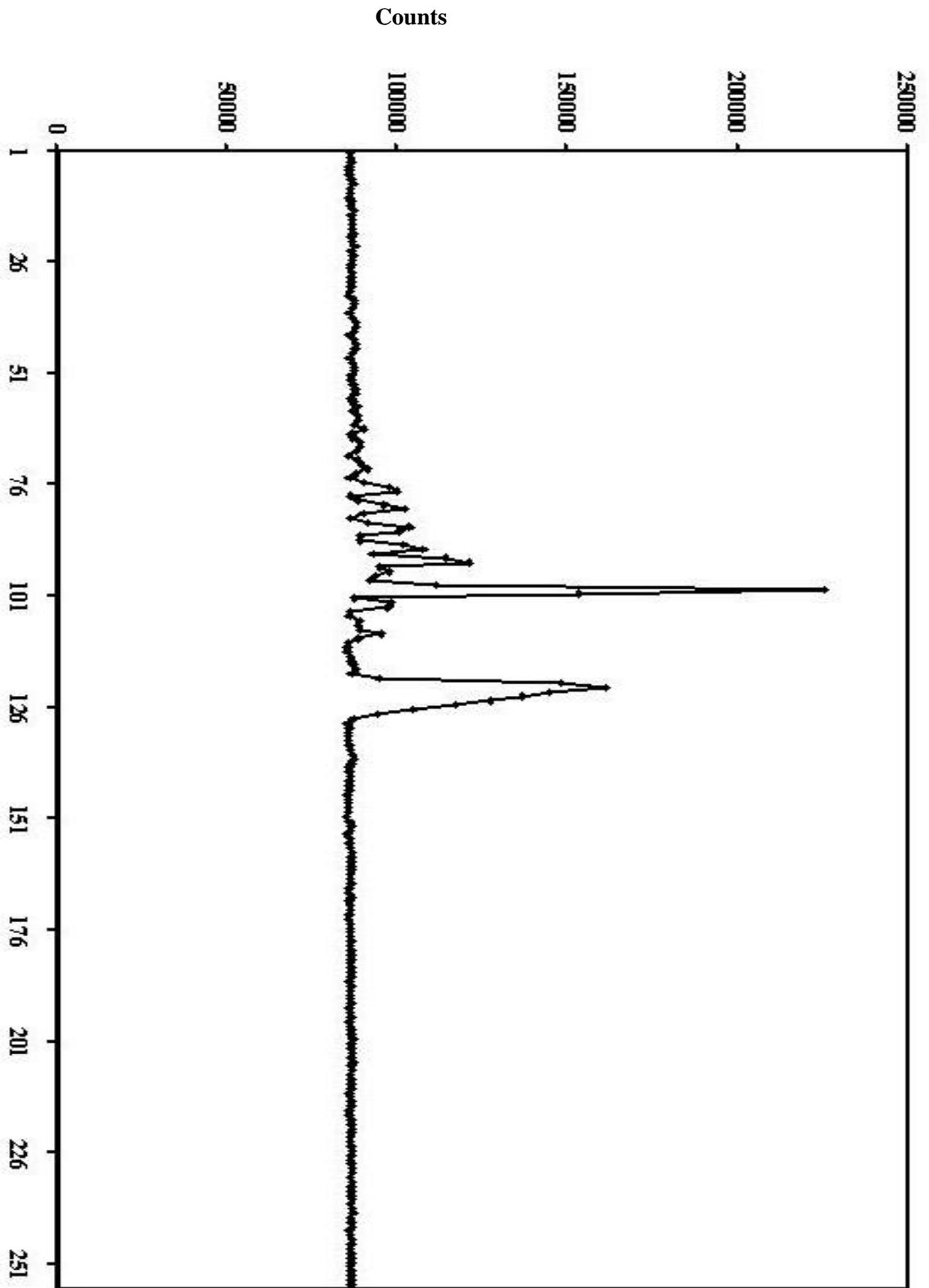

**Fig. 5**

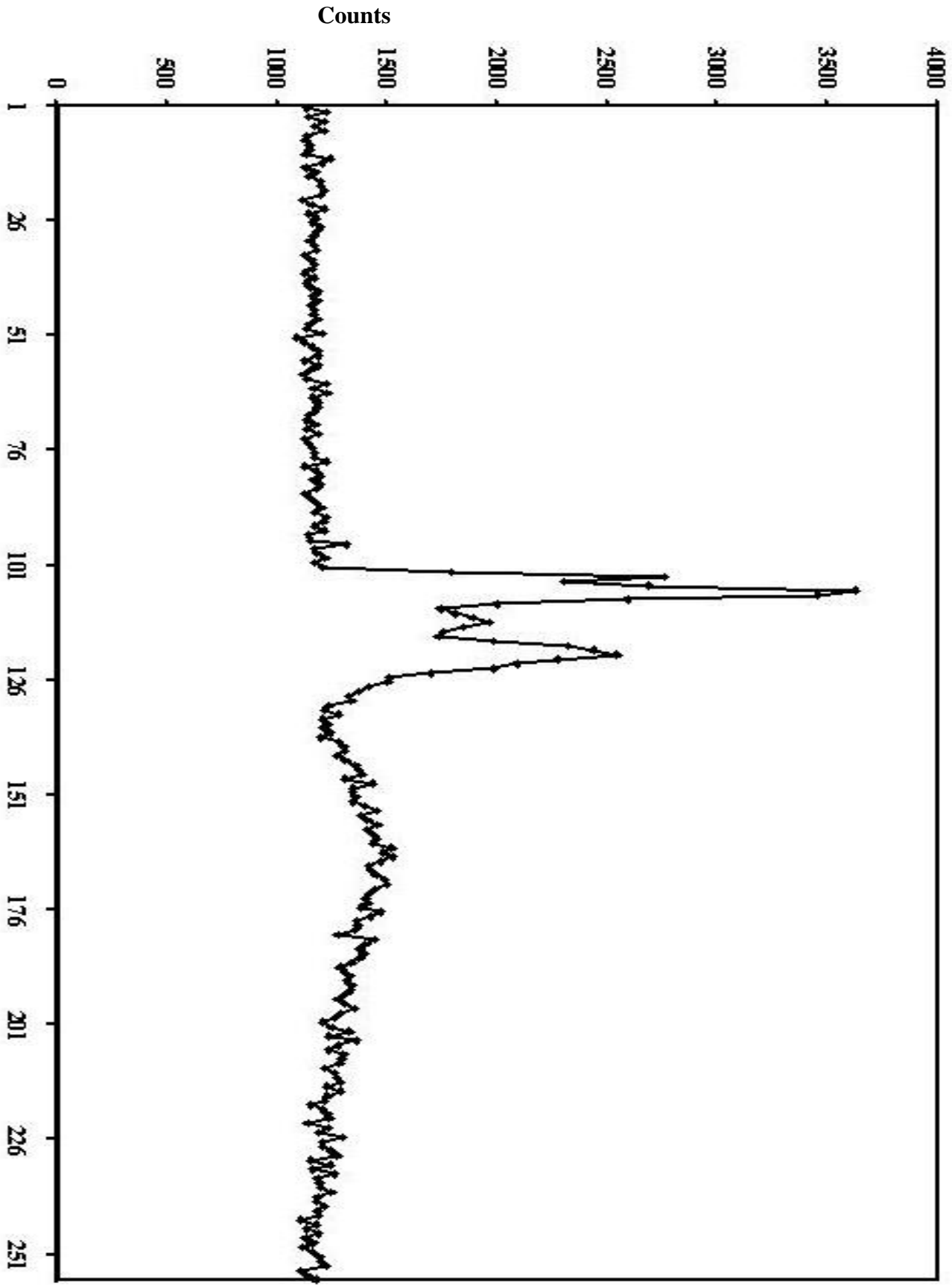

**Fig. 6**